# Generalized Wall-functions for High-Reynolds-number Turbulence Models

*S.V.Utyuzhnikov*

Department of Mechanical, Aerospace & Manufacturing Engineering,
UMIST, PO Box 88, Manchester, M60 1QD, UK

Department of Computational Mathematics,
Moscow Institute of Physics & Technology, Dolgoprudny 141700, Russia

s.utyuzhnikov@umist.ac.uk

**Abstract**

Generalized wall-functions in application to high-Reynolds-number turbulence models are derived. The wall-functions are based on transfer of a boundary condition from a wall to some intermediate boundary near the wall (usually the first nearest to a wall mesh point but that is not obligatory). The boundary conditions on the intermediate boundary are of Robin-type and represented in a differential form. The wall-functions are obtained in an analytical easy-to-implement form, take into account source terms such as pressure gradient and buoyancy forces, and do not include free parameters. The log-profile assumption is not used in this approach. Both Dirichlet and Newman boundary-value problems are considered. A method for complementing solution near a wall is suggested. Although the generalized wall-functions are realized for the *k*-ε model, generalization to other turbulence models looks quite clear. The general approach suggested is applicable to studying high-temperature regimes with variable laminar viscosity and density. A robust numerical algorithm is proposed for implementation of Robin-type wall-functions. Preliminary test results made for a channel flow showed good accuracy and a weak dependence of the solution on the location of the intermediate boundary where the boundary conditions are set.

# 1. Introduction

The problem of mathematical simulation of turbulent flows near walls appears in many practical applications. It is well known that turbulence vanishes near a wall due to the no-slip boundary condition for the velocity as well as the blocking effect caused by the wall. In the vicinity of the wall, there is a thin sublayer with predominantly molecular diffusion. The sublayer has a substantial influence upon the remaining part of the flow. An adequate numerical resolution of a solution in the sublayer requires a very fine mesh because of sublayer thinness and high gradients of the solution. It makes a model used time consuming and often not suitable for industrial applications. Because of low velocities in the sublayer, models that resolve the sublayer are called low-Reynolds-number (LR) models.

To avoid direct resolution of the viscous sublayer, so-called high-Reynolds-number (HR) models have been developed. In this type of models the sublayer domain is not directly resolved. It significantly saves computational efforts [1]. In the HR models, the boundary conditions or near-wall profiles are represented by wall-functions. In most cases, the wall-functions are semi-empirical and have very limited applications [1-5]. First wall-functions are based on the log-law profile for the velocity [4, 5]. The main disadvantage of these wall-functions is a strong dependence on the location of the mesh point closest to a wall where the wall-functions are applied. Such a problem is especially pronounced if the first mesh point is located inside the viscous sublayer. To avoid this, the scalable wall-function approach is suggested in [6]. Wilcox showed [7] that pressure gradient must be taken into account to avoid the mesh dependence. In more recent approaches [1-3] attempts have been made to take into account the pressure gradient and other effects such as buoyancy forces. Numerical comparisons done in [1-3] showed that such advanced wall-functions give substantially better prediction than the standard wall-functions. A brief review of different wall-functions used can be found in, e.g., [1]. Sub-grid numerical wall-functions are developed in [2] where dependent variables are determined by solving boundary-layer-type transport equations in a near-wall sub-grid. In this approach, the boundary condition on the boundary that is external to the wall is determined by linear interpolation of certain main-grid values. In [3], the analytical wall-functions are evolved by integrating boundary-layer-type equations in the vicinity of a wall using the assumption that all terms besides the dissipative one are constant. Mainly, it means that the contribution of the convective terms is neglected near the wall and the pressure gradient and buoyancy force (if applicable) are not changed. At the wall, the boundary conditions are the same as those used in the LR models. The analytical profiles are then used in the cell nearest to the wall to reconstruct the near-wall solution. The wall-functions for the turbulent kinetic energy and its dissipation are based on a local approximation in the near-wall cell and assumption on the local equilibrium. Although approaches [1-3] allows one make substantially better prediction in comparison to the standard methods, the realization of them is quite complicated especially in case of their implementation into industrial codes. The wall-functions [1-3] can be represented in a finite-difference form only. Although this form is suitable for finite-volume algorithms, its use for finite-difference approximations is not clear. As the standard wall-functions, this approach faces substantial problems if the nearest to a wall cell is in the viscous sublayer. Also, it is important to note that the second to the wall cell can not be much smaller or bigger than the first one because of integration over the first cell used.

In the following sections, the method of boundary condition transfer is used [8]. The method allows us to transfer a boundary condition from a wall to some intermediate surface. It will be showed that it is possible to transfer a boundary condition either approximately (analytically) or exactly (numerically). The boundary conditions on the intermediate surface are always of Robin-type and represented in a differential form. They can take into account the influence of the source terms in governing equations. These boundary conditions are interpreted as generalized wall-functions. The location of the point, where the boundary conditions transfer to, does not make any considerable influence on a mesh distribution nearby this point. The wall-functions can easy be used especially in finite-difference approximations. A method for complementing the solution on the entire domain outside the viscous sublayer is suggested.

Numerical intermediate Robin-type boundary conditions are developed along with a decomposition method. The method allows us to split the boundary-value problem into two parts: an inner near-wall (internal) problem and an outer (external) one. Both boundary-value problems can be solved independently, using different numerical schemes and meshes, which yields the terminal solution. The inner solution can be then used for complementing the solution near the wall if HR models are applied along with the generalized wall-functions.

The general approach suggested is applicable to studying high-temperature turbulent flows with variable density and laminar viscosity.

## 2. Model equation

First, considering the following model equation:

$$(\mu u_y)_y = R_h, \qquad (1)$$

defined in a domain $\Omega = [0, y_e]$ with Dirichlet boundary conditions:

$$u(0) = u_0, \ u(y_e) = u_1 \qquad (2)$$

Assuming that $R_h = const$, after integrating equation (1) from $0$ to $y$, one obtains:

$$\tau_w = \mu(y)du/dy(y) - R_h\, y, \qquad (3)$$

where $\tau_w = \mu(y)du/dy(y)_w$. Index $w$ here and below means a value at $y = 0$. The second integration gives

$$u(y) = u_0 + \tau_w \int_0^y \frac{1}{\mu} d\xi + R_h \int_0^y \frac{\xi}{\mu} d\xi \qquad (4)$$

Considering (3) and (4) at some intermediate point $y^*$ in $\Omega$ and excluding $\tau_w$, we have:

$$u(y^*) = u_0 + f_1 \frac{du}{dy}(y^*) - \frac{R_h}{\mu^*} f_2, \qquad (5)$$

where

$$f_1 = \int_0^{y^*} \frac{\mu(y^*)}{\mu(y)} dy, \quad f_2 = \int_0^{y^*} \frac{\mu(y^*)}{\mu(y)}(y^* - y) dy \qquad (6)$$

Relation (5) can be considered as a Robin-type boundary condition transferred from a wall ($y = 0$) to some point $y^*$ inside the domain $\Omega$. This boundary condition can be either exact (if the exact function of $\mu$ is used in (6)) or approximate (if $\mu$ is estimated by one way or another).

If $R_h = R_h(y)$, then

$$u(y^*) = u_0 + f_1 \frac{du}{dy}(y^*) - \left(\int_0^{y^*} R_h dy\right) \frac{f_2}{y^* \mu^*}, \qquad (7)$$

where

$$f_1 = \int_0^{y^*} \frac{\mu(y^*)}{\mu(y)} dy, \quad f_2 = y^* \int_0^{y^*} \frac{\mu(y^*)}{\mu(y)} \left(1 - \int_0^{y} R_h dy' / \int_0^{y^*} R_h dy\right) dy \qquad (8)$$

Assuming that the coefficient $\mu$ varies piece-wise linearly

$$\mu = \begin{cases} \mu_w & \text{if } 0 \leq y \leq y_v \\ \dfrac{y - y_v}{y^* - y_v}(\mu^* - \mu_w) + \mu_w & \text{if } y_v \leq y \leq y^*, \end{cases}$$

it is possible to obtain analytical expressions for $f_1$ and $f_2$ if $R_h = const$ and $y_v \leq y^*$:

$$f_1 = \alpha_\mu y_v (1 + \theta \ln \alpha_\mu), \quad f_2 = \alpha_\mu y_v \left[(1-\theta)y^* + (\theta^2 \alpha_\mu \ln \alpha_\mu - 1/2 + \theta)y_v\right], \qquad (9)$$

where $\alpha_\mu = \mu^*/\mu_w$, $\theta^{-1} = \dfrac{\mu^* - \mu_w}{\mu_w} \dfrac{y_v}{y^* - y_v}$. The parameter $\theta$ represents cotangent of the inclination angle of the dependence $\mu/\mu_w$ on $y/y_v$.

In a more general case, it is not difficult to take into account variation of the coefficient $\mu$ in the interval $[0, y_v]$ as in [3].

Once a Newman problem is solved:

$$du/dy(0) = \tau_w/\mu_w, \quad u(y_e) = u_1, \qquad (10)$$

the intermediate boundary condition at $y = y^*$ is simpler than in the previous case:

$$\frac{du}{dy}(y^*) = (\tau_w + \int_0^{y^*} R_h dy)/\mu^* \tag{11}$$

The boundary conditions are always linear if governing equation (1) is linear. They are represented in a differential-integral form which can be easy realized especially with finite-difference approximations. It is easy to see that mesh distribution possible in the interval $[y^*, y_e]$ can be independent of the location of the intermediate boundary corresponding to the point $y^*$.

In application to the Reynolds averaged Navier-Stokes equations (RANS), the method of boundary condition transfer considered here gives generalized Robin-type wall-functions.

## 3. Generalized wall functions

We apply the method of boundary condition transfer given above to derive the generalized wall-functions for the tangential velocity component $U$ or temperature $T$, and the turbulent kinetic energy $k$.

Neglecting diffusion parallel to a wall, the momentum and enthalpy transport equations can be written in the Cartesian coordinate system $(x, y)$ as follows:

$$\frac{\partial}{\partial y}\left[(\mu_l + \mu_t)\frac{\partial U}{\partial y}\right] = \rho U \frac{\partial U}{\partial x} + \rho V \frac{\partial U}{\partial y} + \frac{\partial P}{\partial x} + \beta(T)g \tag{12}$$

$$\frac{\partial}{\partial y}\left[\left(\frac{\mu_l}{Pr} + \frac{\mu_t}{Pr_t}\right)\frac{\partial T}{\partial y}\right] = \rho U \frac{\partial T}{\partial x} + \rho V \frac{\partial T}{\partial y} \tag{13}$$

Here $\mu_l$ and $\mu_t$ are the laminar and turbulent viscosities, accordingly; $Pr$ and $Pr_t$ are Prandtl numbers; $U$ and $V$ are the velocity component in the $(x, y)$ coordinate system; $\rho$ is the density; $P$ is the pressure; $\beta(T)g$ is the term representing the buoyancy effect if applicable.

Both equations have the same form as model equation (1) where the right-hand side $R_h$ equals either $\rho U \frac{\partial U}{\partial x} + \rho V \frac{\partial U}{\partial y} + \frac{\partial P}{\partial x} + \beta(T)g$ or $\rho U \frac{\partial T}{\partial x} + \rho V \frac{\partial T}{\partial y}$.

Then, the intermediate boundary conditions for $U$ and $T$ at point $y^*$ are given by (7) upon substitution either $U$ or $T$ instead of $u$ accordingly. Evidently, the coefficient $\mu$ must be considered as either $\mu_l + \mu_t$ or $\mu_l/Pr + \mu_t/Pr_t$. In case of the momentum equation $u_0 = 0$. If $y^*$ is chosen in the vicinity of the wall, the right-hand side $R_h$ can be represented by a constant neglecting the terms with y-derivatives and taking the terms with x-derivatives at $y^*$, as it is made in [3]. Thus, in case of the enthalpy and momentum equations the relative right-hand sides are as follows:

$$R_h = R_{ht} \equiv \rho U dT/dx(y^*), \qquad (14)$$

$$R_h = R_{hu} \equiv \rho U dU/dx(y^*) + dP/dx(y^*) + \beta(T(y^*)) \qquad (15)$$

Unlike [3], a similar approach is applied to the equation for the turbulence kinetic energy as well:

$$\frac{\partial}{\partial y}\left[\left(\mu_l + \frac{\mu_t}{Pr_k}\right)\frac{\partial k}{\partial y}\right] = \rho U \frac{\partial k}{\partial x} + \rho V \frac{\partial k}{\partial y} - P_k + \rho\varepsilon, \qquad (16)$$

where $P_k$ is the production of the turbulent kinetic energy, $\varepsilon$ is its dissipation; $Pr_k$ is the Prandtl number.

Neglecting the x-derivatives, one obtains the following evaluation for $R_h$ in form (7) applicable to the k-equation:

$$R_h(y) = R_{hk} \equiv \rho \frac{dk}{dx}(y^*) + \rho\varepsilon - \mu_t\left(\frac{dU}{dy}\right)^2 \qquad (17)$$

Assuming a piece-wise linear behaviour of the function $\mu_t$ as in [3]:

$$\mu_t = \begin{cases} 0 & \text{if } y < y_v \\ \frac{y - y_v}{y^* - y_v}\mu_t^* & \text{if } y_v < y < y^* \end{cases}, \qquad (18)$$

where $y_v$ is the thickness of the viscous sublayer near the wall. Then, the coefficients $f_1$ and $f_2$ in (8) (the latter term only if $R_h = const$) can be evaluated from (9) where

$$\alpha_\mu = \mu^*/\mu_l, \quad \theta = \frac{y^* - y_v}{y_v} \frac{\mu_l}{\mu_t^*}, \quad \mu^* = \mu_l + \mu_t^* \tag{19}$$

If the turbulent viscosity $\mu_t^*$ in (18) is evaluated as follows [3]:

$$\mu_t = C_\mu C_l \rho \frac{\sqrt{k^*}}{\mu_l} y_v \frac{y^* - y_v}{y_v} \mu_l = C_\mu C_l \tilde{y}_v \frac{y^* - y_v}{y_v} \mu_l \approx 2.5 \frac{y^* - y_v}{y_v} \mu_l, \tag{20}$$

where $k^* = k(y^*)$; $C_\mu = 0.09$, $C_l = 2.55$, $\tilde{y}_v \equiv \rho \frac{\sqrt{k^*}}{\mu_l} y_v = 10.8$, then $\theta$ is a constant equalled to 0.4.

It has been obtained by computations that it is more accurate to evaluate the turbulent viscosity at $y^*$ from the HR $k$-$\varepsilon$ model directly

$$\mu_t^* = C_\mu \rho (k^*)^2 / \varepsilon^* \tag{21}$$

rather than from equation (20).

The sublayer thickness $y_v$ is evaluated as follows [3]:

$$y_v = 10.8 \mu_l / (\rho \sqrt{k^*}). \tag{22}$$

It is assumed that the value $k^*$ corresponds to the fully turbulent region, so it does not tend to zero.

If $y^* < y_v$ then the boundary conditions are set inside the sublayer, and formulas (9) are not valid. The consideration of the boundary conditions (7), (8) inside the sublayer is not appropriate because the model used does not take into account the blocking wall effects adequately. It is suggested to pose the boundary conditions at the edge of the sublayer $y = y_v$ as in [6] because $y_v$ is small enough. Then, the coefficients $f_1$ and $f_2$ in (7) can be evaluated as follows:

$$f_1 = \alpha_\mu y_v, \quad f_2 = \alpha_\mu y_v^2 / 2 \tag{23}$$

It is assumed that the turbulent viscosity $\mu_t$ reaches value (21) at the edge of the viscous sublayer immediately. These boundary conditions are consistent with boundary conditions (9) taking in the limit $y^* \to y_v$ or $\theta \to 0$.

To evaluate $\varepsilon^*$, in [3] it is supposed that in the vicinity of a wall $\varepsilon(y)$ is a continuous function which is constant near the wall and have an inverse dependence on $y$ further. It gives the following approximation for $\varepsilon(y)$:

$$\varepsilon(y) = \begin{cases} \dfrac{(k^*)^{3/2}}{C_l y_d} & \text{if } y < y_d \\ \dfrac{(k^*)^{3/2}}{C_l y} & \text{else} \end{cases}, \qquad (24)$$

where $y_d = 2 C_l \mu_l /(\rho \sqrt{k^*})$. Dependence (24) assumes a variation of $\varepsilon$ in the viscous sublayer because $y_d \approx 0.5 y_v$.

The wall-function for the turbulent energy $k$ is used in form (7), (8) and depends on the right-hand side $R_{hk}(y)$ represented by equality (17). It includes the dissipation $\varepsilon$ and derivative $dU/dy$. The former term is taken from (24) while the last term is evaluated in the interval $[0, y^*]$ from the momentum equation:

$$(\mu_t + \mu_l) dU/dy = \mu^* dU/dy(y^*) + (y - y^*) R_{hu} \qquad (25)$$

In (25), the turbulent viscosity $\mu_t$ is defined by (18).

Thus, the coefficients $f_1$ and $f_2$ in the wall-functions (5)-(9) depend on $y^*$ and $k^*$ only. The latter value is determined from solution of the HR model at the boundary point $y^*$. Hence, the intermediate boundary conditions (7) at $y = y^*$ complete the boundary-value problem in the interval $[y^*, y_e]$ ($y_e$ is the external boundary of the computational domain) and can be considered as generalized wall-functions. These boundary conditions are of Robin-type and similar to the "slip boundary condition" at the edge of the Knudsen-layer in aerodynamics. It is important to note that the boundary conditions are linear with respect to the leading variable. As it follows from (5) and (7), the source terms in the wall-functions can only be essential far enough from a wall because of the quadratic term in $y^*$.

We note that, although the problem is solved in the bulk domain $[y^*, y_e]$ only, the flux to the wall (e.g., skin friction) can be easy evaluated considering (25) (or its analogy for the temperature in case of heat flux) at $y = 0$.

Upon obtaining a HR solution, it can be extended to the interval $[y_v, y^*]$ using the analytical solution if $y_v \leq y^*$:

$$u(y) = u(0) + \varphi_1(y)u_y^* - \varphi_2(y)\frac{R_h}{\mu^*}, \tag{26}$$

$$\varphi_1 = \alpha_\mu y_v (1 + \theta \ln \Omega(y)),$$

$$\varphi_2 = \alpha_\mu y_v \left[ y^* - \theta y + (\theta^2 \alpha_\mu \ln \alpha_\mu - 1/2 + \theta)y_v \right],$$

$$\Omega = 1 + (\alpha_\mu - 1)\frac{y - y_v}{y^* - y_v},$$

It means that the intermediate boundary is not obligatory to be related to the nearest to the wall cell. It is possible to take $y^*$ far enough from the wall and complement the solution on the region to the sublayer by (26).

If a heat transfer problem is considered, where buoyancy force is important, it is not difficult to take into account variable function $\beta(T(y))$. In this case, the analytical solution (26) for the temperature (if $y_v \leq y \leq y^*$) is substituted to the right-hand side $R_h$ for the momentum equation. Since $R_h$ is variable, the wall-functions for the velocity are used in form (7), (8). If $y^*$ is in the sublayer, then $\beta$ can be sufficiently evaluated by the value $\beta(T_w)$. Also, dependence $\mu_l = \mu_l(T)$ can be taken into account using either linear or quadratic approximation, as in [3], and easily implemented.

The generalized wall-functions obtained and their realization are not based on a numerical integration in the inner region $[0, y^*]$, as in [1-3], therefore the location of the intermediate boundary is very substantial for mesh distribution in the bulk domain. It means we can choose, e.g., a fine mesh despite a relatively big value $y^*$ (or vice versa) without loose of stability.

## 4. Numerical solution

In numerical simulation of turbulence, numerical schemes, which reserve positiveness of a solution, are very efficient because unknown variables such as the turbulent kinetic energy $k$ or its dissipation $\varepsilon$ must be positive. The following numerical procedure has been developed for using the positive type schemes in solving boundary-value problems with Robin-type boundary conditions.

Boundary condition (7) can be rewritten in the following general form:

$$k(0) = \alpha \, dk/dx(0) + \beta, \tag{27}$$

assuming that both function $k$ and its derivative $dk/dx$ are positive. This assumption is valid in case of real physical problems for the turbulent kinetic energy in the vicinity of a wall. The coefficient $\alpha$ is positive because $f_1$ is always positive but the coefficient $\beta$ can be negative. In computations it can lead to a negative value of $k$. To avoid such a case, it is suggested to rewrite (27) in the following form if $\beta < 0$:

$$k(0) = \alpha \, dk/dx(0) + \beta \frac{k(0)}{k^-(0)},$$

or

$$k(0) = \alpha' \, dk/dx(0), \qquad (28)$$

where $\alpha' = \dfrac{\alpha}{1 - \beta/k^-(0)} > 0$, and $k^-(0)$ is the value of $k(0)$ taken from the previous time step or iteration.

For stability, near the points, where $\beta$ changes its sign, a relaxation procedure for the coefficients $\alpha$ and $\beta$ has been used.

## 5. Decomposition method

In this section a decomposition method for solving equations in the LR models is derived. The main idea is given below for an arbitrary linear differential equation.

First, considering a linear Dirichlet problem in the interval $[0, y_e]$:

$$Lu = f \qquad (29)$$
$$u(0) = u_0, \qquad u(y_e) = u_1.$$

The entire computational domain $\Omega = [0, y_e]$ is decomposed by two sub-domains, an inner one $\Omega_1 = [0, y^*]$ and outer one $\Omega_2 = [y^*, y_e]$, where $y^* < y_e$.

Near the wall (in the inner domain $\Omega_1$), the following two boundary-value problems are solved:

$$Lu_1 = f, \quad u_1(0) = u_0, \; du_1/dy\,(y^*) = 0 \qquad 0 \le y \le y^*, \qquad (30)$$

$$Lu_2 = 0, \quad u_2(0) = 0, \; du_2/dy(y^*) = 1 \qquad 0 \le y \le y^*. \qquad (31)$$

It is easy to prove that the general solution to (29) on the inner domain $\Omega_1$ is

$$u(y) = u_1(y) + du/dy(y^*)u_2(y) \qquad (32)$$

If we consider (32) at the point $y^*$, we have a Robin-type boundary condition for the outer domain $\Omega_2$:

$$u(y^*) = u_1(y^*) + du/dy(y^*) \, u_2(y^*) \qquad (33)$$

This boundary condition is exact if we set it at $y = y^*$. Thus, the boundary condition from the wall ($y = 0$) is transferred to the point $y^*$.

The problem on one domain $\Omega$ is split into two problems on the domains $\Omega_1$ and $\Omega_2$. As a result, we have some version of a decomposition method.

In case of Newman boundary conditions

$$du/dy(0) = u_0, \qquad u(y_e) = u_1 \qquad (34)$$

the algorithm is similar to that for the Dirichlet problem (29). Indeed, we solve the following two boundary-value problems:

$$Lu_1 = f, \quad du_1/dy(0) = u_0, \quad u_1(y^*) = 0, \qquad (35)$$

$$Lu_2 = 0, \quad du_2/dy(0) = 0, \quad u_2(y^*) = 1. \qquad (36)$$

The general solution to problem (34) on the inner domain $\Omega_1$ is

$$u(y) = u_1(y) + u(y^*) u_2(y) \qquad (37)$$

After derivation, a Robin-type boundary condition at $y^*$ is obtained:

$$du/dy(y^*) = du_1/dy(y^*) + u(y^*) du_2/dy(y^*) \qquad (38)$$

After solution to problems (35) and (36), we use this boundary condition in the outer domain $\Omega_2$. In the inner domain $\Omega_1$, the solution to problem (34) is then obtained from (37), since $u(y^*)$ is known from the outer problem on $\Omega_2$.

In case of nonlinear equations, the decomposition procedure given above is used in nonlinear iterations.

Thus, the decomposition method described above allows us to split the problem into the two parts: the near-wall (including the viscous sublayer) problem and the outer one. At a first glance, we gain nothing obtaining three boundary-value problems instead of one. On the other hand, in the inner (near-wall) and outer domains the appropriate problems can be solved on different meshes using different approximations. The analysis of LR models shows that the behaviour of the solution in the vicinity of the sublayer defers drastically from that in the rest bulk domain, therefore such a splitting can be useful. It is important to note that in uniform approaches (one domain only) the presence of two adjacent cells having substantially different sizes can lead to a loss of accuracy or even stability. Additionally, one can note that the algorithms on solving the inner and outer problems can be easily parallelized.

Since the functions $u_1(y)$ and $u_2(y)$ do not explicitly depend on the solution in the domain $\Omega_2$ (the dependence is via the coefficients only), these functions can be calculated once and then used further along with the use of either HR or LR model in an external domain (similar to $\Omega_2$). In this approach, the problem is only solved in the external domain, and the solution is complemented on the entire domain then by the following way:

$$v(y) = \alpha u_1(y) + u_y^* u_2(y), \tag{39}$$

$$\alpha = \frac{u^* - u_2^* u_y^*}{u_1^*}$$

It is easy to verify that $v(y^*) = u^*$, $dv/dy(y^*) = u_y^*$. It guarantees a smooth junction of both the inner and outer solutions.

To determine approximate "basic" functions $u_1$ and $u_2$, the coefficients in the relative problems can be evaluated using Reichardt's profiles for all the variables $u$, $k$ and $\varepsilon$ based on some estimation for $u_\tau$.

This method can be effective if similar boundary-value problems are considered. For instance, if an initial-boundary value 2D problem is studied, we solve a boundary value problem in one direction $x$ and an initial problem in another direction $y$. It is possible to calculate the "basic" functions $u_1$ and $u_2$ once at some point $x_0$ and use them for approximate complementing the solution further at the next points $x_i$.

The decomposition approach, as well as the wall-functions developed, can be easy generalized and used for the problems with variable viscosity and density.

## 6. Test case

A channel fully developed plane flow has been conducted as a test case. The flow is simulated far enough from the edge of the channel, so that the problem can be considered as 1D [9]. The standard HR $k$-$\varepsilon$ model has been used to test the wall-function approach:

$$\frac{\partial}{\partial y}\left[(\nu + \nu_t)\frac{\partial U}{\partial y}\right] = p_x/\rho,$$

$$\nu_t\left(\frac{\partial U}{\partial y}\right)^2 - \varepsilon + \frac{\partial}{\partial y}\left[(\nu + \nu_t/\Pr_k)\frac{\partial k}{\partial y}\right] = 0,$$

$$C_{\varepsilon 1}\frac{\varepsilon}{k}\nu_t\left(\frac{\partial U}{\partial y}\right)^2 - C_{\varepsilon 2}\frac{\varepsilon^2}{k} + \frac{\partial}{\partial y}\left[(\nu + \nu_t/\Pr_\varepsilon)\frac{\partial \varepsilon}{\partial y}\right] = 0, \qquad (40)$$

$$\nu_t = C_\mu k^2/\varepsilon,$$

$$C_{\varepsilon 1} = 1.44, C_{\varepsilon 2} = 1.92, \Pr_k = 1, \Pr_\varepsilon = 1.3$$

Here $y$ is the distance to the wall, $p_x$ is the pressure gradient in the channel which is assumed to be negative, and $\nu = \mu/\rho$.

In the computations given below the Reynolds number is $Re \equiv u_\tau h/\nu = 395$, where $u_\tau = \sqrt{-hp_x/\rho}$ is the friction velocity, $h$ is the half of the channel height. The dependence of dimensionless velocity, $u^+ = U/u_\tau$, on the universal coordinate, $y^+ = yu_\tau/\nu$, will be calculated using the approach developed in this work and compared against the benchmark results.

As it follows from Section 4, the generalized wall-functions have only one parameter which is the coordinate $y^*$ of the point where the boundary conditions are moved to. In this work, calculations have been done for different values of $y^*$ including those less than $y_\nu$ (in the viscous sublayer). In Figure 1 the velocity profiles $u^+$ obtained by the wall-function approach are given against $y^+ = yu_\tau/\nu$ for different values of $y^*$. The profiles are compared against Reichardt's profile [10] representing the benchmark solution.

In Figure 1 the velocity profile is given for $y^{+*} \equiv \dfrac{y^* u_\tau}{\nu} = 1; 5; 10$. Although $y^*$ is deeply in the viscous sublayer the correspondence to the LR solution (Reichardt's profile) is quite reasonable. The maximal difference falls at the case $y+^* = 10$ when the point $y^*$ is nearby (but outside) the sublayer. A slight difference between the first two curves ($y+^*$ equals 1 and 5) is explained by different boundary conditions for $\varepsilon$.

If $y^*$ is outside the sublayer, the prediction is better. It is shown by Figure 2 where $y+^* = 30; 50; 100; 200$. Even, if $y^*$ is taken far from the viscous sublayer, the solution is quite close to the benchmark solution. In the last case ($y^* = 200$) a half of the domain is excluded in the case of the HR model; nevertheless, in the bulk (rest) domain the solution is quite sensible.

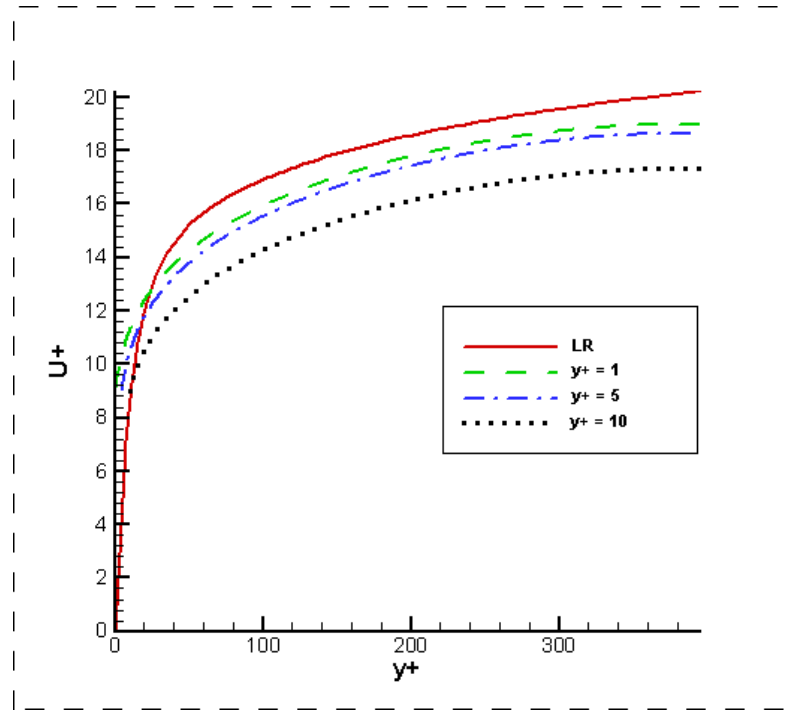

Fig. 1. Velocity profile in channel flow. Solid line is Reichardt's profile; the other lines correspond to $y+^* = 1; 5; 10$.

In all range of $y^*$ considered ($1 \leq y+^* \leq 200$) the difference in $u^+$ predicted by the LR profile and HR models is within 15%.

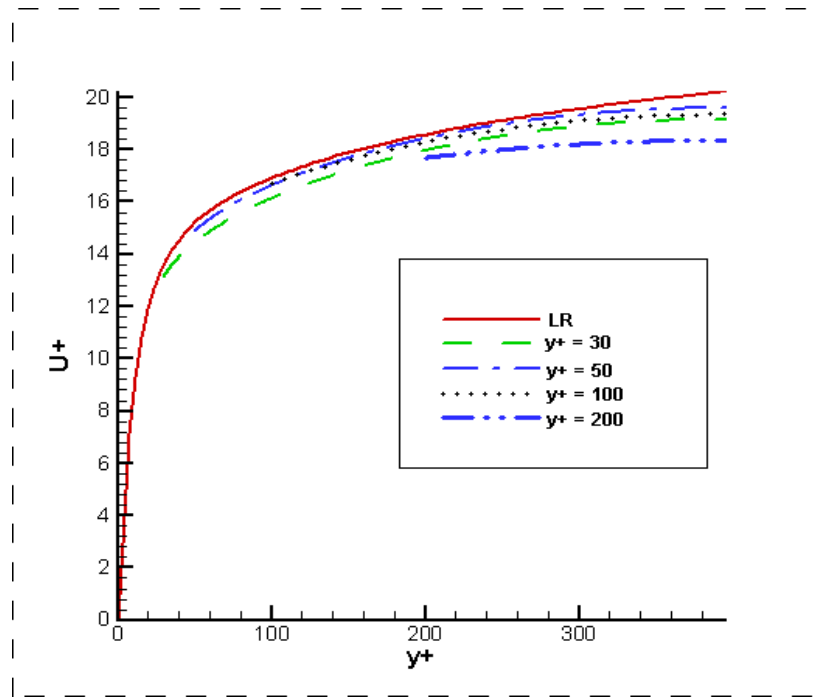

Fig. 2. Velocity profile in channel flow. Solid line is Reichardt's profile; the other lines correspond to $y+^* = 30; 50; 100; 200$.

A comparison between several different kinds of wall-functions is shown in Figure 3 for $y^{+*} = 100$. The dashed line corresponds to the homogeneous boundary conditions where the right-hand side is not taken into account. In this case, $f_2 = 0$ in either (5) or (7). The error is big enough; that confirms the importance of taking into consideration the source terms in the wall-functions in the case of big values of $y^*$. If the source term is included in the boundary conditions for the velocity only, the prediction then becomes much more accurate (dash-dotted line). At the same time, the replacement of the boundary condition (7) for the turbulent kinetic energy $k$ by a frequently used boundary condition $k = 1/3 u_\tau^2$ gives a substantial deviation from the benchmark solution (dotted line). The standard wall-functions for all variables [9]

$$u^* = (1/\kappa \ln(u_\tau y^*/\nu) + 5) u_\tau^2, \quad k^* = 1/3 u_\tau^2, \quad \varepsilon^* = \kappa u_\tau^3 / y^* \qquad (41)$$
$$\kappa = 0.41$$

results in much better prediction (dash-double-dotted line). Nevertheless, one can note that the channel flow is one of the most convenient test cases for the standard wall-functions. The current approach gives more accurate prediction, shown in Figure 2, although the main advantages of the wall-functions developed can be expected at consideration of more complicated cases. It is important to emphasise that in contrast to the standard wall-functions the approach suggested is not based on the assumption on the velocity log-profile or any similar additional information to match the solution. Also, we note that the standard wall-functions include the friction velocity $u_\tau$, which is uknown in advance in a general case, that makes the boundary conditions nonlinear and demands appropriate iterations for resolving.

The decomposition method described in Section 5 has been tested using the low-Reynolds number model by Chien [11]:

$$\nu_t \left(\frac{\partial U}{\partial y}\right)^2 - \varepsilon + \frac{\partial}{\partial y}\left[(\nu + \nu_t / \Pr_k)\frac{\partial k}{\partial y}\right] = 0$$

$$C_{\varepsilon 1} f_1 \frac{\tilde{\varepsilon}}{k} \nu_t \left(\frac{\partial U}{\partial y}\right)^2 - C_{\varepsilon 2} f_2 \frac{\tilde{\varepsilon}^2}{k} + E + \frac{\partial}{\partial y}\left[(\nu + \nu_t / \Pr_\varepsilon)\frac{\partial \tilde{\varepsilon}}{\partial y}\right] = 0 \qquad (42)$$

$$\varepsilon = \varepsilon_0 + \tilde{\varepsilon}, \quad \nu_t = C_\mu f_\mu k^2 / \tilde{\varepsilon}, \quad f_\mu = 1 - e^{-0.0115 y^+},$$

$$f_1 = 1, f_2 = 1 - 0.22 e^{-(\text{Re}_T / 6)^2}, \quad C_1 = 1.35, \quad C_2 = 1.8,$$

$$\varepsilon_0 = 2\nu \frac{k}{y^2}, \quad E = -2\nu \frac{\tilde{\varepsilon}}{y^2} e^{-y^+/2}, \quad \text{Re}_t = \frac{k^2}{\tilde{\varepsilon} \nu}$$

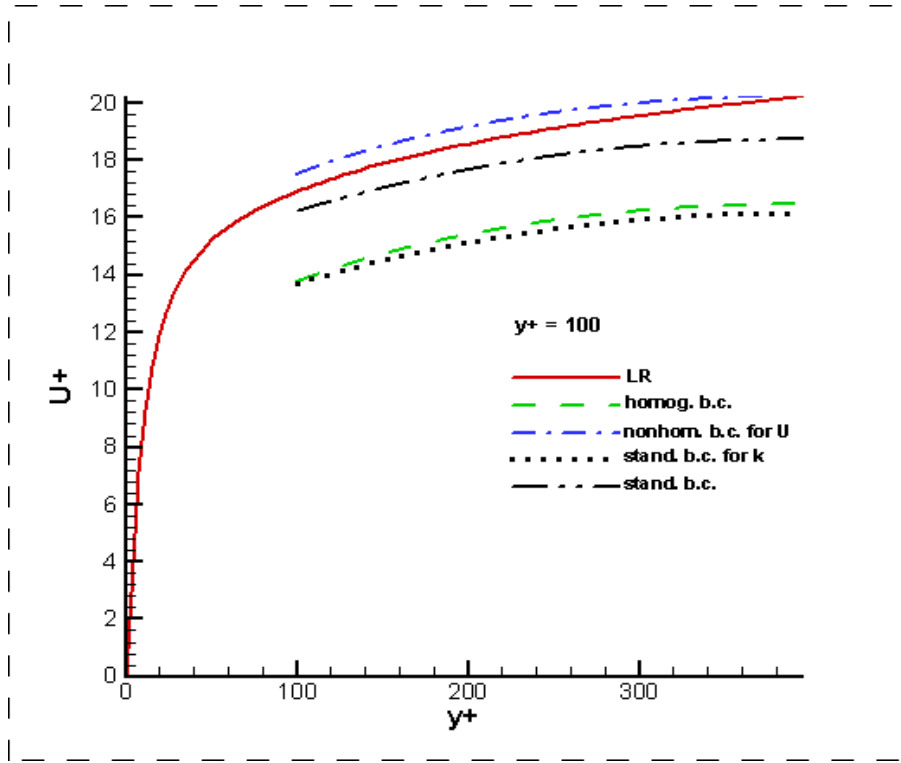

Fig. 3. Velocity profile in channel flow with different wall-functions. Solid line is Reichardt's profile; the other lines correspond to homogeneous wall-functions for $U$ and $k$ (dashed line), and only for $k$ (dashed-dotted line); "standard" wall-function for $k$ (dotted line) and for all variables (dashed-double-dotted line).

In Figure 4 the dashed and dotted lines represent the solution obtained by the decomposition method with the junction point at $y^{+*} = 100$. In each sub-domain $\Omega_1$ and $\Omega_2$ a uniform mesh with 20 points is used. For comparison, the dashed-dotted line corresponding to 1-block solution obtained on a mesh with 40 points is given. The deviation from Reichardt's profile is explained by a very coarse mesh (in case of a fine mesh the prediction is much more accurate). The HR solution complemented on the whole domain by LR solution (39) is shown by the curves with circles and diamonds. The complemented part of this solution on the interval $[0 \; y^{+*}]$ is given in Figure 5 (dashed line). The curves marked by squares and triangles represent the "basic" near-wall solutions $u_1$ and $u_2$ used in (39). The dotted line is the analytical solution (26).

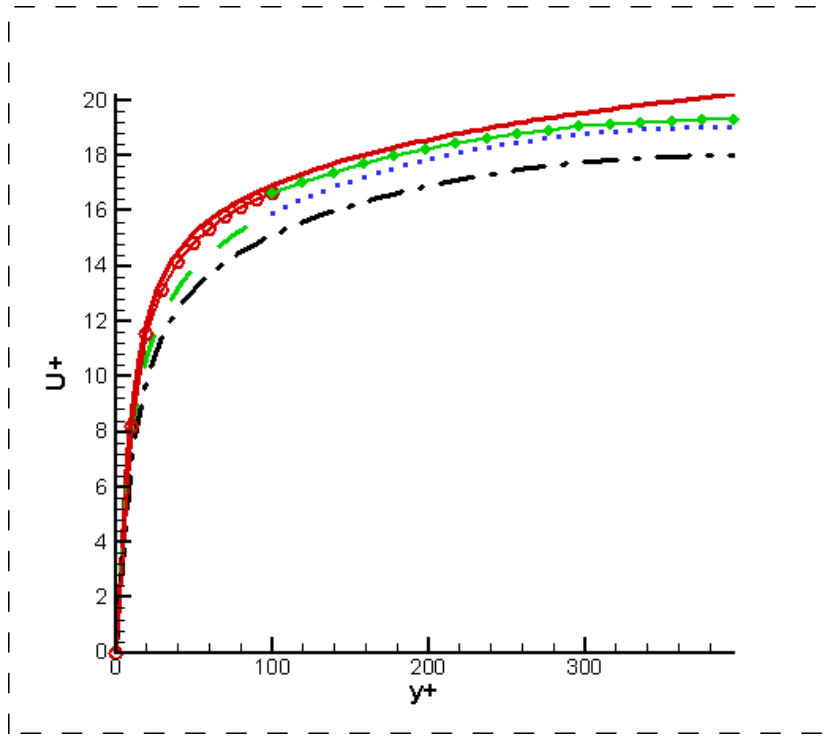

Fig. 4. Velocity profile in channel flow. Solid line is Reichardt's profile; dashed and dotted lines is solution of LR model by decomposition method with 20 and 20 points (each part); dashed-dotted line is 1-block solution with 40 points. Curves with squares and triangles are basic solutions $u_1$ and $u_2$ accordingly. Curve with circles and diamonds is composite LR and HR solution.

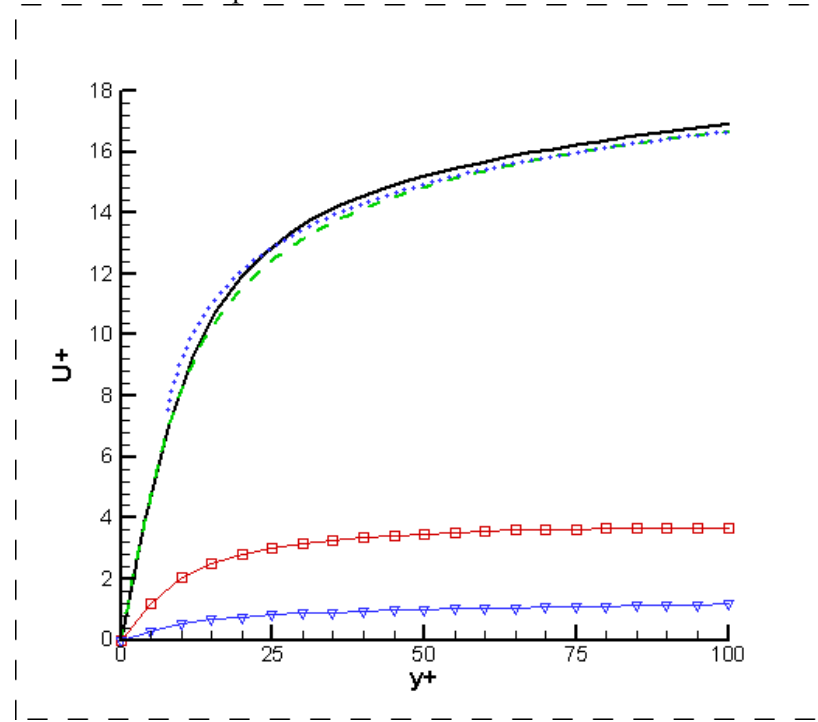

Fig. 5. Complemented velocity profiles. Solid line is Reichardt's profile; dashed is LR complementation (39); dotted line is analytical solution (26). Curves with squares and triangles are basic solutions $u_1$ and $u_2$ accordingly.

# 7. Conclusion

Generalized wall-functions have been developed. They are based on transfer of boundary conditions from a wall to some point in the computational domain (usually the nearest to the wall grid point). The boundary conditions at this point are of Robin-type and represented in a differential form. These boundary conditions are interpreted as generalized wall-functions taking into account source terms and used for HR models. The wall-functions have been obtained in a compact easy-to-implement analytical form and they do not include any adjustable parameters. Testing this approach along with the k-ε equations applied to a fully developed turbulent flow in a channel showed that the proposed wall-functions are quite accurate even if the boundary conditions are set at a point either in a viscous sublayer or far beyond. A numerical robust approach preserving positivity of a solution in the case of Robin-type boundary conditions has been developed. Mesh distribution inside the computational domain can be chosen independently of the location of the intermediate boundary.

In application to LR models a decomposition method has been suggested. It allows us to split the problem into a near-wall part and the rest one. The boundary-value problems in both parts can be solved independently using different numerical schemes and meshes. The near-wall part of the solution can be used for approximate complementing the solution obtained by the wall-functions and HR model near a wall on the entire domain including a near-wall part. This opportunity requires an additional investigation on the base of consideration of multidimensional problems. The decomposition approach, as well as the wall-functions proposed, can be easy generalized and used for the problems with variable viscosity and density.

The generalized wall-functions suggested performed well in a 1D test problem on channel fully turbulent flow. The wall-functions can be relatively easy extended on a multidimensional case. In the future, a quantitative study for such problems is expected.

# 8. Acknowledgment

The author is grateful to A.V. Gerasimov and D.R.Laurence for fruitful discussions, S.A. Snegirev for useful remarks.

This work has been supported by the FLOMANIA project (Flow Physics Modelling - An Integrated Approach) is a collaboration between Alenia, AEA, Bombardier, Dassault, EADS-CASA, EADS-Military Aircraft, EDF, NUMECA, DLR, FOI, IMFT, ONERA, Chalmers University, Imperial College, TU Berlin, UMIST and St. Petersburg State Technical University. The project is funded by the European Union and administrated by the CEC, Research Directorate-General, Growth Programme, under Contract No. G4RD-CT2001-00613.